\documentclass[12pt,a4paper]{article}
\pdfoutput=1
\usepackage[utf8x]{inputenc}
\usepackage{ifthen}
\usepackage{amssymb}
\usepackage{amsmath}
\usepackage{graphicx}
\usepackage{hyperref}

\usepackage{color}

\newcommand{\eq}{\begin{equation}}
\newcommand{\eqx}{\end{equation}}
\newcommand{\eqs}{\begin{equation*}}
\newcommand{\eqsx}{\end{equation*}}
\newcommand{\eqn}{\begin{eqnarray}}
\newcommand{\eqnx}{\end{eqnarray}}
\newcommand{\eqns}{\begin{eqnarray*}}
\newcommand{\eqnsx}{\end{eqnarray*}}
\newcommand{\alg}{\begin{align}}
\newcommand{\algx}{\end{align}}

\newcommand{\f}[2]{\frac{#1}{#2}}

\newcommand{\cor}[1]{\left\langle{#1}\right\rangle}

\newcommand{\lm}{\lambda}

\renewcommand{\th}{\theta}
\newcommand{\sg}{\sigma}
\newcommand{\Sg}{\Sigma}

\newcommand{\om}{\omega}
\newcommand{\Om}{\Omega}
\newcommand{\gm}{\gamma}

\newcommand{\qqqq}{\quad\quad\quad\quad}
\newcommand{\tr}{\mbox{\rm tr}\,}
\newcommand{\CC}{{\mathbb{C}}}
\newcommand{\nn}{{\cal N}}

\newcommand{\dw}{\partial}
\newcommand{\dwb}{\bar{\partial}}
\newcommand{\wb}{\bar{w}}
\newcommand{\jb}{\bar{j}}
\newcommand{\Jb}{\bar{J}}
\newcommand{\Psihat}{\hat{\Psi}}

\newcommand{\infp}{\infty^+}
\newcommand{\infm}{\infty^-}
\newcommand{\zerp}{0^+}
\newcommand{\zerm}{0^-}
\newcommand{\plone}{I_+}
\newcommand{\mnone}{I_-}
\newcommand{\sqj}{{\sqrt{1+j^2}}\,}
\newcommand{\dlogth}{\phi}
\newcommand{\red}[1]{{#1}}
\newcommand{\violet}[1]{{#1}}

\newcommand{\UP}{UP}
\newcommand{\DN}{DN}

\newcommand{\smallhalf}{\textstyle\frac12}

\DeclareMathOperator \cn {cn}
\DeclareMathOperator \dn {dn}
\DeclareMathOperator \sn {sn}
\DeclareMathOperator \am {am}
\DeclareMathOperator \diag {diag}

\newcommand{\vc}[2]{\left(
\begin{matrix}
{#1} \\
{#2} \\
\end{matrix}
\right)}

\newcommand{\arr}[4]{\left(
\begin{matrix}
{#1} & {#2}\\
{#3} & {#4} \\
\end{matrix}
\right)}

\title{Surprises in the AdS algebraic curve constructions
--- Wilson loops and correlation functions}

\author{Romuald A. Janik${}^{\,a,b,}$\thanks{e-mail: {\tt romuald@th.if.uj.edu.pl}},\ 
Paweł Laskoś-Grabowski${}^{\,a,c,}$\thanks{e-mail: {\tt plg@th.if.uj.edu.pl}}}

\date{\vspace{6pt}${}^a$ Institute of Physics, Jagiellonian University\\
ul. Reymonta 4, 30-059 Kraków, Poland\\
\vspace{6pt}${}^b$ Institute for Advanced Studies\\
The Hebrew University of Jerusalem\\ 
Givat Ram Campus, 91904 Jerusalem, Israel\\
\vspace{6pt}$^c$ Institute for Theoretical Physics, University of Wrocław\\
pl. Maxa Borna 9, 50-204 Wrocław, Poland\\}

\begin{document}

\maketitle

\begin{abstract}
The algebraic curve (finite-gap) classification of rotating string
solutions was very important in the development of integrability
through comparison with analogous structures at weak coupling.
The classification was based on the analysis of monodromy around 
the closed string cylinder.
In this paper we show that certain classical Wilson loop minimal
surfaces corresponding to the null cusp and $q\bar{q}$ potential
with \emph{trivial monodromy} can, nevertheless, be described by
appropriate algebraic curves. We also show how a correlation
function of a circular Wilson loop with a local operator fits
into this framework. The latter solution has identical monodromy
to the pointlike BMN string and yet is significantly different.
\end{abstract}

\vfill{}

\section{Introduction}

The AdS/CFT correspondence, which postulates the equivalence
of $\nn=4$ Super-Yang-Mills theory and superstrings in $AdS_5 \times
S^5$ spacetime, provides a unique opportunity for solving,
for the first time, an interacting four-dimensional gauge theory
(see the recent review \cite{INTREVIEW}).

Currently we have a very good understanding of the spectral problem,
i.e.\ of anomalous dimensions of local gauge theory operators
at any coupling, which gets translated to the 
energy levels of the closed string in $AdS_5 \times S^5$
spacetime, i.e.\ the energy levels of the corresponding
worldsheet quantum field theory. The answer is formulated
in terms of Thermodynamic Bethe Ansatz \cite{TBA1,TBA2,TBA3}
or NLIE \cite{NLIE1,NLIE2} equations for that integrable
worldsheet QFT. Of course, there are still many issues which
have not been worked out, like the structure of source terms
in these equations for arbitrary operators, but still
our understanding is much more complete than for other
observables.

A significant step in the above progress was the development
of the algebraic curve (finite-gap) classification of
classical spinning string solutions \cite{KMMZ,KZ,BKSZ,ALGREVIEW}
and a comparison of the emerging structures with similar
classical analysis of the Bethe equations coming from
a gauge theory spin chain description at weak coupling.

Because of this theoretical importance, our motivation
was to investigate whether one could employ similar
algebraic curve methods for other classes of classical
string solutions in $AdS_5 \times S^5$ which also have
an important gauge-theoretical meaning -- Wilson loops
and correlation functions.

Since the method of \cite{KMMZ} was based on a thorough
analysis of the analytical properties of monodromy
around a noncontractible loop going around the closed
string cylinder, it would seem that there is no chance
of applying similar constructions to Wilson loop minimal
surfaces on which all loops are contractible and hence
have trivial monodromy. Indeed, the use of integrability
for polygonal null Wilson loops related to scattering
amplitudes \cite{AM,AGM,YW} relied on completely
different methods specific to that particular setup.

In this paper we will show that, nevertheless, one can
associate algebraic curves to such classical solutions
as the null cusp minimal surface or the $q\bar{q}$
potential Wilson loop and conversely, one can
reconstruct the full target space solution purely
algebraically from the given algebraic curve.

The consideration of classical solutions corresponding
to correlation functions of local operators \cite{JSW,BT,JW,BT2}
(and possibly also other objects like Wilson loops \cite{CORRLOOP}) 
poses a different kind of question to the classical algebraic curve 
construction of \cite{KMMZ}. For these solutions, the monodromy
around a given puncture should, \emph{by definition}, be identical
to the monodromy (pseudomomentum) of the spinning string
corresponding to the local operator at the puncture.
Therefore, the starting point of the construction of \cite{KMMZ}
would be identical for the ordinary spinning string and for the correlation
function. Yet the classical solutions are significantly 
different. This shows that there should be an enormous freedom
in the construction of solutions with prescribed pseudomomentum,
going far beyond the folklore that such solutions are parametrized
essentially by a finite-dimensional Jacobian of the relevant
algebraic curve.
The understanding of how this freedom may arise was one of the 
motivations for this paper. We investigate here the correlation
function of a circular Wilson loop with the $\tr Z^J$
local operator and show how the algebraic curve description
differs from the pointlike string corresponding to $\tr Z^J$.

Finally, another motivation for describing Wilson loops in the
same setting as closed strings was to understand from this
perspective possible links between the two quite different kinds
of solutions. An outstanding example of such a relationship
is the link between the large spin limit of the GKP folded
string \cite{GKP} and the null cusp Wilson loop \cite{KRUCZGKP}. This
relationship allowed for the identification of the cusp
anomalous dimension appearing in the Wilson loop
with the large spin asymptotics of the anomalous
dimensions of twist-two operators. Hence one could use the
well developed Bethe ansatz methods (in this limit wrapping
corrections do not contribute) at any coupling (for the spinning string)
to gain all-order information on the Wilson loop. 

Having a unified description of both kinds of solutions may help
understanding such relationships and perhaps uncover new ones.
One of the motivations for the present work was to investigate
the possibility of such a relationship with the $q\bar{q}$
Wilson loop minimal surface. The simplest deformation yields
unfortunately just the generalized $q\bar{q}$ Wilson loop of 
\cite{FD}, yet perhaps
there might be a more intricate generalization through e.g.\ 
a degeneration of a genus-2 curve. 

Let us note, that as this paper was being prepared, significant
progress was made in the exact evaluation of the $q\bar{q}$
potential \cite{CHMSqq,Drukkerqq}. It would be very interesting
to understand whether the algebraic curve for the $q\bar{q}$
potential identified here has any interpretation in these
approaches.

The plan of the paper is as follows. In section~2 we review briefly
the classical integrability of the $AdS_3$ $\sg$-model, and in section~3,
the algebraic curve (finite-gap) description of spinning strings and
the reconstruction procedure which enables one to obtain the target-space
solution from the algebraic curve. In section~4 we summarize the key
questions of the present paper. In section~5 we present our
example classical solutions and identify the corresponding algebraic curves,
while in the following section, we show that one can indeed reconstruct
the original Wilson loop minimal surface or correlation function solution
just from the knowledge of the algebraic curves and some minimal 
structural assumptions. We close the paper with conclusions and some remarks 
on possible applications. In the appendices, we have collected
some formulas for elliptic functions and we identify the
classical solutions arising from algebraic curves approximating 
the null cusp (here we obtain the GKP folded string) and the 
$q\bar{q}$ potential (in this case we obtain the known generalized
minimal surface in global $AdS_3$).

\section{The $AdS_3$ $\sg$-model and its integrability}

It is very well known that the full $AdS_5 \times S^5$ $\sg$-model is 
integrable \cite{BPR}. 
In this paper, for simplicity, we will concentrate on its subsector, 
the $AdS_3$ $\sg$-model, which is also classically integrable by itself. 
In order to exhibit integrability, it is most convenient to rewrite its action
in terms of group elements:
\eq
S_{AdS_3}=\f{\sqrt{\lm}}{4\pi} \int \tr j \jb\, d^2 w
\eqx
where $w$ and $\wb$ are the worldsheet coordinates,\footnote{We may take these
coordinates to be either complex or light-cone depending on the worldsheet signature.}
the currents are given by
\eq
j=g^{-1} \dw g \qqqq \jb=g^{-1} \dwb g
\eqx
with the group element having one of the following three forms
\eq
\label{e.group}
g=\arr{\f{ix_1+x_2}{z}}{\f{1}{z}}{-\f{x_1^2+x_2^2+z^2}{z}}{\f{ix_1-x_2}{z}}
,
\arr{\f{x_1+x_2}{z}}{\f{1}{z}}{-\f{-x_1^2+x_2^2+z^2}{z}}{\f{x_1-x_2}{z}}
,\arr{e^{i t} \cosh \rho}{e^{i \psi} \sinh \rho}{e^{-i\psi} \sinh \rho}{e^{-i t} \cosh \rho}
\eqx
depending on whether we are considering Euclidean $AdS_3$, Minkowskian $AdS_3$ in the
Poincaré patch, or global $AdS_3$ respectively.

Integrability of the $AdS_3$ $\sg$-model means that there is a family
of flat currents parametrized by an arbitrary complex number --- the spectral 
parameter $x\in \CC$. Namely defining
\eq
J=\f{j}{1-x} \qqqq \Jb=\f{\jb}{1+x}
\eqx
we find that the equations of motion are equivalent to the flatness condition
enforced for arbitrary $x$:
\eq
\dw \Jb- \dwb J +[J,\Jb]=0
\eqx
For the following, it will be important to regard the above flatness 
condition
as the compatibility condition for the auxiliary linear problem
\begin{align}
\label{e.linear}
\dw \Psi+J \Psi  &= 0 \nonumber\\
\dwb \Psi+\Jb \Psi  &= 0
\end{align}
where $\Psi(w,\wb;x)$ is a 2-component vector. Once one 
knows two independent solutions of (\ref{e.linear}), one can put them into
two columns of a $2\times 2$ matrix $\Psihat(w,\wb;x)$ which satisfies
the matrix differential equations
\eq
\label{e.linearhat}
\dw \Psihat+J\Psihat=0 \qqqq \dwb \Psihat+\Jb\Psihat=0
\eqx
The knowledge of $\Psihat(w,\wb;x=0)$ allows us to reconstruct the original
string classical solution. Namely, we can at once get the currents from
\eq
\label{e.jpsihat}
j=-\dw \Psihat \cdot \Psihat^{-1} |_{x=0}
\eqx
as well as reconstruct the classical solution by the formula \cite{DV}
\eq
\label{e.gpsihat}
g=\sqrt{\det \Psihat}\cdot  \Psihat^{-1} |_{x=0}
\eqx
It would seem at first glance that these formulas are not particularly 
useful, since in order to find $\Psihat$ directly one would have to solve the
system (\ref{e.linearhat}) which depends on the knowledge of the classical
solution (which is encoded in the currents $J$, $\Jb$). The algebraic curve
construction (or `finite-gap construction') allows, however, 
to construct $\Psihat(w,\wb;x)$
directly starting from a given algebraic curve and exploiting general
analyticity properties of its dependence on the spectral parameter $x$.
This procedure is described in general
in Chapter 5 of \cite{BABELON}
and in the context of spinning strings in $AdS$ in \cite{DV,DVPHD}.
As we will be using it in a quite general form in the present paper,
we will review it below.

\section{A brief review of the finite gap (algebraic curve) construction}
\label{s.finitegap}

In the context of strings in $AdS_5 \times S^5$, the algebraic curve construction
has been adopted exclusively for the case of spinning string solutions.
These are classical, closed string solutions of the relevant bosonic $\sg$-model.

The starting point of the construction is the monodromy
operator associated to the flat currents defined above.
\eq
\label{e.monodromy}
\Om(w_0,\wb_0;x)= P e^{\int_C J dw+\Jb d\wb}
\eqx
where $(w_0,\wb_0)$ is some reference point on the worldsheet, and $C$ is
a curve going from this point once around the cylinder and going back
to $(w_0,\wb_0)$. The flatness of the currents implies that the monodromy 
does not depend on smooth deformations of $C$ (hence, if $C$ were contractible,
the resulting monodromy would be trivial). An easy consequence of 
this path
independence is the behaviour of the monodromy w.r.t. a change of 
the reference point:
\eq
\Om(w_1,\wb_1;x)= U\, \Om(w_0,\wb_0;x)\, U^{-1}
\eqx
Hence the eigenvalues of the monodromy operator do not depend on the 
reference 
point and thus are conserved (they may be computed e.g.\ using
any constant time circle on the worldsheet cylinder). 
These eigenvalues depend on the spectral parameter
and so this construction provides an infinite set of conserved quantities.
For the case at hand, the eigenvalues can be written as
\eq
e^{ip(x)} , \, e^{-ip(x)}
\eqx
where $p(x)$ is the so-called pseudomomentum. The crucial input for the algebraic
curve classification of the finite-gap solutions are the analytic properties
of $p(x)$ as a function of the spectral parameter.

Let us note at this stage, that it would seem that the whole algebraic curve
method would be inapplicable for Wilson loop solutions, for which all loops
are contractible and hence the monodromy is trivial -- so there is no
pseudomomentum to start with. In this paper we will show that in fact one can
adopt the algebraic curve classification method to Wilson loops and we will
show explicitly how one can associate algebraic curves to certain standard
examples and, conversely, how one can explicitly reconstruct these solutions
algebraically starting from the given algebraic curves.

An algebraic curve can be constructed out of the monodromy operator
by defining
\eq
L(w,\wb;x)=-i \f{\partial}{\partial x} \log \Om(w,\wb;x)
\eqx
which is a $2\times 2$ matrix with rational coefficients and then
defining
\eq
\det(\tilde{y} \cdot 1-L(w,\wb;x))=0
\eqx
which clearly only depends on $p'(x)$. Redefining $\tilde{y}$ to get rid of
double poles at $x=\pm 1$ (see \cite{DV} for a discussion) gives the standard
genus $g$ algebraic curve $\Sg$ of the form
\eq
y^2=\prod_{i=1}^{2g+2} (x-a_i)
\eqx 
together with a meromorphic differential form $dp$ on $\Sg$, which satisfies
a set of conditions (see \cite{KMMZ,KZ,BKSZ,ALGREVIEW}) allowing e.g.\ for 
computing the energy of the classical solution in terms of conserved charges 
(spins, angular momenta). We are not presenting these expressions here,
as we will not use them in the following.

Let us mention a subtlety associated with the algebraic curve of the
$AdS_3$ $\sg$-model. For the commonly studied case of $AdS_3 \times S^1$ \cite{KZ},
the pseudomomentum form $dp$ has double poles at $x=\pm 1$ associated
to the $J$ charge of the $S^1$:
\eq
p(x) \sim \f{\pi \f{J}{\sqrt{\lm}}}{x \pm 1}+\ldots
\eqx
When $J=0$ and the solution is completely contained in $AdS_3$ (i.e.\ satisfies
Virasoro constraints there), $x= \pm 1$ become \emph{branch points} of the algebraic
curve $\Sg$ and e.g.\ for the GKP folded string solution at $J=0$, $dp$ has
the form
\eq
dp =\f{A x^2+B}{(x^2-1) \sqrt{(x^2-1)(x^2-a^2)}}dx
\eqx
with an algebraic curve $y^2=(x^2-1)(x^2-a^2)$. See appendix~\ref{s.gkp}
for a discussion of the reconstruction of the GKP solution from
this algebraic curve.

\subsection*{Reconstruction of the classical solutions from
algebraic curves}

Let us now briefly sketch how to reconstruct the full classical solution
from an algebraic curve for the spinning string introduced above.
We will review the reconstruction procedure specializing initially to
the spinning string context and indicating, at the end, the
passage to the most general case of \cite{BABELON}.

The monodromy operator (\ref{e.monodromy}) is just the parallel 
transport \emph{of the solutions
of the linear system} (\ref{e.linear}) around a cycle which
goes once around the worldsheet cylinder. So given a fixed point 
$(w,\wb)$ on the worldsheet we will have two distinguished solutions 
of (\ref{e.linear}) which will
be the eigenvectors of $\Om(w,\wb;x)$ corresponding to $e^{\pm i p(x)}$. 
The algebraic curve $\Sg$ can be understood as encoding the information 
how these solutions depend on the spectral parameter $x$ (keeping
the reference point $(w,\wb)$ fixed). In particular the two branches of
the curve above $x$ correspond to these two solutions. More precisely, both
solutions, as functions of $x$, can be described by a \emph{single} 
vector-valued function on $\Sg$. By abuse of notation, we will write
$x\in \Sg$ when we mean either of the two points in $\Sg$ lying above $x$.
These two points will be denoted explicitly by $x^+$ and $x^-$.

Reconstruction starts from the realization that $\Om$ can be simultaneously
diagonalized with the linear operators $\dw+J$ and $\dwb+\Jb$.
Hence a solution of (\ref{e.linear}) should be proportional to the
eigenvector of $\Om$, the determination of which is a somewhat simpler
problem.
The eigenvector of $\Om$ can be normalized as
\eq
\Om \Psi_n(w,\wb;x)=e^{ip(x)} \Psi_n(w,\wb;x)
\quad \text{with} \quad
\Psi_n(w,\wb,x)=\vc{1}{\psi_n(w,\wb;x)}
\eqx
which defines a single scalar function $\psi_n(w,\wb;x)$.
$\psi_n(w,\wb;x)$ is a meromorphic function 
on $\Sg$ and, for a genus $g$ curve $\Sg$, \emph{typically}
has $g+1$ poles.\footnote{See Proposition on p. 133 of \cite{BABELON},
which rests, however, on some
genericity assumptions. We will encounter later an important example
where this is violated.} Moreover $g$ of these poles will move on $\Sg$ 
as we change the worldsheet point $(w,\wb)$. These are called \emph{dynamical
poles}. 
In order to proceed further, one writes the most general 
form of $\psi_n(w,\wb;x)$ consistent with these assumptions.

In the second step of the reconstruction procedure, we use the fact 
that a solution of (\ref{e.linear})
should be proportional to the above eigenvector:
\eq
\Psi(w,\wb;x)=f_{BA}(w,\wb;x) \cdot \Psi_n(w,\wb;x)
\eqx
The function $f_{BA}(w,\wb;x)$ is called a Baker-Akhiezer function on $\Sg$
(treated as a function of $x$). It has to satisfy certain analyticity
conditions, in particular it should
\begin{enumerate}
\item vanish at the dynamical poles
\item have an essential singularity of a prescribed form at the special
points $x=\pm 1$ (see \cite{DV})
\eq
\label{e.essentialorg}
f_{BA}(w,\wb;x) \sim e^{const \cdot \f{w}{x - 1}} \quad\quad
f_{BA}(w,\wb;x) \sim e^{const \cdot \f{\wb}{x + 1}}
\eqx
\item as $x\to \infty$, $\Psi(w,\wb;x)$ should become independent of the
worldsheet coordinates.  
\end{enumerate}
Let us note that it is only in the last two conditions above, that the 
worldsheet coordinates enter explicitly. Remarkably enough, 
all these conditions are enough to
allow one to reconstruct the full $w,\wb$ and $x$ dependence of the
solution of the linear system (\ref{e.linear}), and thus the original
classical string solution through (\ref{e.jpsihat})-(\ref{e.gpsihat}).

Let us note that the reconstruction procedure, as sketched here
following \cite{BABELON}, does not
really depend too much on the fact that we used the monodromy operator
$\Om(w,\wb;x)$. We could have, and probably should have, used instead its 
logarithmic derivative $L(w,\wb;x)$. 
But in fact, for the whole procedure to work, we also do not need
to use the specific construction of $L(w,\wb;x)$. 
What is enough is that it is a Lax operator, i.e.\ a $2\times 2$ 
matrix satisfying
\begin{align}
\dw L + [J,L] &= 0 \nonumber\\
\dwb L+ [\Jb,L] &= 0
\end{align}
whose entries are \emph{rational} (or polynomial) functions of $x$.
However, once we make this generalization, we will have to generalize
and rederive the condition for the essential singularity 
(\ref{e.essentialorg}) of the Baker-Akhiezer function.

\section{Questions}

After this brief review of the classical algebraic curve approach to
classical spinning string solutions we are ready to formulate the key
questions which were a motivation for this work.

{\bf Question 1.} The classical algebraic curve approach 
in the $AdS_5 \times S^5$
context has been applied for spinning strings, where it started from
the notion of the pseudomomentum associated to the monodromy along 
noncontractible loops. We would like to ask whether one can
adapt this framework to describe the classical solutions
associated to Wilson loop expectation values. For these solutions
there are (typically) no noncontractible loops, so the starting
point of the preceding construction does not even exist.

{\bf Question 2.} Recently, a quite different family of classical 
string solutions, began to be considered. These are classical solutions
corresponding to multi-point correlation functions of $\nn=4$ SYM
operators associated with classical spinning strings.
These solutions have the topology of a punctured sphere and the external
states may be identified with concrete classical spinning strings
by requiring that the monodromy around a given puncture coincides
\emph{exactly} with the monodromy (pseudomomentum) $p(x)$ of a
given classical spinning string solution.
This obvious fact is very surprising taking into account the folklore
that the space of string solutions with a given pseudomomentum
is finite dimensional (e.g.\ these solutions should be
parametrized just by $g$ positions of the dynamical poles 
plus some finite data etc.).
However since the monodromy around each puncture 
is characterized \emph{exactly} by 
the pseudomomentum $p_i(x)$ of the corresponding spinning string, 
the classical solution associated to  
a correlation function with this given operator will also be described
by \emph{the same} algebraic curve as the corresponding finite-gap
spinning string solution.
Hence the class of solutions
with a given monodromy should be much richer than naively expected.
How is this possible? Even more so, there should exist classical
string solutions which should be simultaneously associated with
three or more distinct algebraic curves! We will not attempt here
to address this problem in full generality, but rather study
a correlation function of a Wilson loop with a local operator
which exhibits similar phenomena (namely identical pseudomomentum
with the original spinning string).


{\bf The strategy.} 
Recall that the algebraic curve construction for a curve of genus $g$
implies a very particular dependence of the solution of the linear
system (\ref{e.linear}) as a function of the spectral parameter $x$.
Our approach to the above questions is to study, for some 
specific examples, 
the associated solutions of (\ref{e.linear})
and see whether the analytic structure of these explicit solutions 
implies the existence of a hidden algebraic curve.
We will do it for the null cusp and $q\bar{q}$ 
potential Wilson loop minimal surfaces and for a correlator
of a circular Wilson loop with a local operator.
We will identify the relevant algebraic curves and show
that, based on this information alone, one may reconstruct the
original Wilson loop solutions.

\section{Examples}

In this section we will introduce our basic examples: the light-like
cusp Wilson loop solution, the $q\bar{q}$ potential minimal surface
and the correlation function of a circular Wilson loop with the
$\tr Z^J$ operator. In each case we will explicitly write the classical
solution, give two independent solutions of the associated linear
system (\ref{e.linear}), and identify an associated algebraic curve
through an explicit construction of a \emph{polynomial} Lax matrix.
In the following section we will show how starting just with that
algebraic curve we may reconstruct the original Wilson loop/correlation
function solution.

\subsection*{The null cusp Wilson loop}

The null cusp is an Euclidean minimal surface embedded in the 
Poincaré patch of Minkowskian $AdS_3$. It is given explicitly 
as \cite{KRUCZGKP,CUSPSOL}
\begin{align}
\label{e.cusp}
t &= e^{-\sqrt{2} \sg} \cosh \sqrt{2}\tau \nonumber\\
x &= -e^{-\sqrt{2} \sg} \sinh \sqrt{2}\tau \nonumber\\
z &= \sqrt{2} e^{-\sqrt{2} \sg}
\end{align}  
with $\tau$, $\sg$ coordinates related to the $w$, $\wb$ ones through
$w=\sg+i\tau$, $\wb=\sg-i\tau$. The solution is defined on the  
whole complex plane, so all loops are contractible.
The minimal surface approaches the boundary when $\sg \to+\infty$.
Then the two null lines forming the cusp are obtained when 
one simultaneously takes $\tau \to \pm \infty$.

The two independent solutions of the linear problem (\ref{e.linear})
take the following explicit form
\eq
\label{e.cusp1}
\Psi_1(w,\wb;x)= e^{-\f{1-i}{4}\sqrt{2} \left( iw \sqrt{\f{1+x}{1-x}}+
\wb \sqrt{\f{1-x}{1+x}} \right)}
\vc{e^{\f{1+i}{4}\sqrt{2} \left(-iw+\wb \right)}}{
e^{-\f{1+i}{4}\sqrt{2} \left(-iw+\wb \right)} (-ix+\sqrt{1-x^2})}
\eqx
and
\eq
\label{e.cusp2}
\Psi_2(w,\wb;x)= e^{\f{1-i}{4}\sqrt{2} \left( iw \sqrt{\f{1+x}{1-x}}+
\wb \sqrt{\f{1-x}{1+x}} \right)}
\vc{e^{\f{1+i}{4}\sqrt{2} \left(-iw+\wb \right)}}{
e^{-\f{1+i}{4}\sqrt{2} \left(-iw+\wb \right)} (-ix-\sqrt{1-x^2})}
\eqx
These solutions have at least part of the structure which is reminiscent
of an underlying algebraic curve. The $\sqrt{1-x^2}$ makes a prominent
appearance, the two solutions differ by choosing a different branch
of the square root, which, as mentioned in section \ref{s.finitegap}, 
is characteristic
of treating the linear solution as a single function on the two
branches of the algebraic curve. Finally the exponential prefactor
is suggestive of a Baker-Akhiezer origin, although its singularity
does not look at first glance as an isolated essential singularity.

In order to unambiguously associate an algebraic curve with this solution,
we will find a \emph{polynomial} Lax matrix $L(w,\wb,x)$, 
i.e.\ a $2\times 2$ matrix
with polynomial dependence on the spectral parameter $x$,
satisfying
\eq
\dw L + [J,L] = 0 \qqqq \dwb L+ [\Jb,L] = 0
\eqx
It is clear that we can solve the above equation by taking \emph{any}
expression of the form
\eq
L(w,\wb,x)=\Psihat(w,\wb;x) \cdot A(x) \cdot \Psihat(w,\wb;x)^{-1}
\eqx
where $\Psihat$ is a matrix whose columns are any two independent solutions
of (\ref{e.linear}) and $A(x)$ is an arbitrary $x$-dependent matrix. In 
general the result will not be a polynomial in $x$.
However, for the case at hand, putting $\Psi_1$ and $\Psi_2$ as columns
of $\Psihat$ and taking $A(x)$ to be $A(x)=\sqrt{1-x^2}\diag(1,-1)$ 
gives the following polynomial Lax matrix
\eq
L(w,\wb,x)=\arr{ix}{e^{\f{1+i}{\sqrt{2}} (-iw+\wb)}}{
e^{-\f{1-i}{\sqrt{2}} (w+i\wb)}}{-ix}
\eqx 
Now the algebraic curve is defined by $\det(y-L(w,\wb;x))=0$, which
gives
\eq
\label{e.algccusp}
y^2=1-x^2
\eqx
This is a genus-0 algebraic curve. 
The reader might be worried that by inserting the $\sqrt{1-x^2}$
factor into $A(x)$ we have put in the answer (\ref{e.algccusp})
by hand. This is not so, since the factor $\sqrt{1-x^2}$ was
crucial in order to have a \emph{polynomial} Lax matrix.

The elementary solutions 
(\ref{e.cusp1}) and
(\ref{e.cusp2}) are, by construction, eigenvectors of $L(w,\wb;x)$, i.e.
\eq
L\,\Psi_1=\sqrt{1-x^2}\, \Psi_1 \qqqq L\,\Psi_2=-\sqrt{1-x^2}\, \Psi_2
\eqx
It will be important in the following that the Lax matrix is
diagonal (with distinct eigenvalues) as $x\to \infty$.

In section \ref{s.recon}, we will show that one can explicitly construct $\Psi_{1,2}$,
and hence the original classical solution (\ref{e.cusp}) in 
a completely standard way starting just from the algebraic curve
(\ref{e.algccusp}).

\subsection*{The $q\bar{q}$ potential Wilson loop}

The $q\bar{q}$ Wilson loop minimal surface, introduced in 
\cite{QQMALD,SJREY}, approaches the boundary at two lines
at a spacelike separation $L$. For our purposes,
we will need a conformally flat worldsheet parametrization
which was first obtained in \cite{QQCHR}:
\begin{align}
z &= z_0\cn\sigma \notag\\
x_1\equiv t &= z_0\tau/\sqrt 2 \notag\\
x_2\equiv x &= z_0F(\sigma)/\sqrt 2
\end{align} 
where
\begin{equation}
F(\sigma) = 2E(\am\sigma|\smallhalf)-\sigma,
\end{equation}
$E$ is the incomplete elliptic integral of the second kind,
\begin{equation}
z_0 = \frac{\Gamma(\frac14)^2}{(2\pi)^{\frac32}}L
\end{equation}
is the maximum bulk extension of the surface (attained at $\sigma=0$), and the
Jacobi amplitude $\am$ and Jacobi elliptic functions $\cn,\sn,\dn$ are always
taken with a parameter $\frac12$, i.e.\ $\am\sigma\equiv\am(\sigma|\frac12)$
etc.\ (for more information on these, see appendix \ref{app.elliptic}). $w,\wb$
are defined identically as in the case of the null cusp, 
but the solution is defined now only on a strip where $\cn\sigma\geq0$,
i.e.\ $|\sigma|\leq K(\frac12)$. At the ends of this interval the worldsheet
forms the two parallel Wilson lines on the boundary.

Starting from the Euclidean signature form of $g$, one proceeds essentially in
the same fashion as in the case of cusp, albeit with significant computational
complications arising due to the special functions involved. 
A basis of independent solutions can be taken as
\begin{align}
\Psi_1 &= \phantom{\frac{1}{\sqrt x}}\frac{\mathcal{E}_+\sqrt{1-x\cn^2\sigma}}{\cn\sigma} \vc{1}{\frac{\sqrt2x\cn\sigma\sn\sigma\dn\sigma+i\sqrt x\sqrt{1-x^2}\cn^2\sigma}{x\cn^2\sigma-1}-\frac{i\tau+F(\sigma)}{\sqrt2}} \label{e.qqpsi1}\\
\Psi_2 &= \frac{i}{\sqrt x}\frac{\mathcal{E}_-\sqrt{1-x\cn^2\sigma}}{\cn\sigma} \vc{1}{\frac{\sqrt2x\cn\sigma\sn\sigma\dn\sigma-i\sqrt x\sqrt{1-x^2}\cn^2\sigma}{x\cn^2\sigma-1}-\frac{i\tau+F(\sigma)}{\sqrt2}} \label{e.qqpsi2}
\end{align}
where
\begin{equation}
\mathcal{E}_\pm = \exp\frac{-i\sigma+\tau x+i(1+x)\Pi(\frac{x}{x-1};\am\sigma|\frac12)}{\pm\sqrt2\sqrt x\sqrt{1-x^2}}
\end{equation}
(with $\Pi$ being the incomplete elliptic integral of the third kind) are the
Baker-Akhiezer-like prefactors and we also notice that everything that 
(essentially) discerns both solutions are different signs of the square root
terms. The proportionality constant $i/\sqrt x$ is
essential to ensure that in the limit $x\to0$ the matrix $\hat\Psi$ 
will be invertible and its determinant positive.

Let us note that the above solution has one feature which naively
excludes the possibility of an underlying algebraic curve -- the factor
\eq
\sqrt{1-x\cn^2\sigma}
\eqx
This would indicate the existence of a branch cut whose position is 
dependent on the worldsheet coordinate, which is at odds with any 
kind of algebraic curve description. One finds, however, that this
branch cut is cancelled by a corresponding cut in $\mathcal{E}_\pm$.

We then construct a polynomial Lax matrix choosing $\hat\Psi=(\Psi_1\Psi_2)$ and
$A(x)=\sqrt x\sqrt{1-x^2}\cdot\diag(1,-1)$. The result is definitely more
complicated than in the previous case, but the characteristic polynomial is
nonetheless simple and the algebraic curve is in this case defined by
\begin{equation}
y^2=x(1-x^2).\label{e.qqcurve}
\end{equation}
This is an elliptic (genus-1) curve. The eigenvalues associated to $\Psi_{1,2}$
as eigenvectors of $L$ are
\begin{equation}
L\,\Psi_1=\sqrt x\sqrt{1-x^2}\,\Psi_1 \qqqq L\,\Psi_2=-\sqrt x\sqrt{1-x^2}\,\Psi_2.
\end{equation}
A significant complication in the present case is the fact that $x=\infty$
is a branch point of the algebraic curve (\ref{e.qqcurve}). Consequently,
the asymptotics of $L(w,\wb;x)$ as $x\to \infty$ are more subtle:
\eq
L(w,\wb;x) \propto x^2 \arr{0}{0}{1}{0} +\ldots
\eqx

\subsection*{The $\cor{W(C)\tr Z^J}$ correlation function}

Let us now consider a classical solution which
corresponds to a correlation function of a circular Wilson loop
with the local operator $\tr Z^J$ (the BMN vacuum). This example
is interesting for a different reason than the previous two.
Now we have a noncontractible loop and nontrivial monodromy,
but that monodromy is completely determined by the pseudomomentum
of the local operator -- here that of the BMN vacuum namely:
\eq
\label{e.pxBMN}
p(x)= \f{2\pi j x}{x^2-1}
\eqx
where $j=J/\sqrt{\lm}$.
This pseudomomentum is identical to the one in a classical configuration
corresponding to a correlation function of two local operators
\eq
\cor{\tr \bar{Z}^J \tr Z^J}
\eqx
(or its global AdS counterpart -- the standard BMN pointlike string).
What distinguishes these two configurations? Clearly the operator
$\tr Z^J$ may appear in arbitrarily complicated correlation functions,
yet all of them will have exactly the same monodromy (\ref{e.pxBMN}).

We will contrast here the two cases: $\cor{\tr \bar{Z}^J \tr Z^J}$
and $\cor{W(C)\tr Z^J}$. By a special conformal transformation, one 
can always put the coordinate of the insertion point of the local
operator $\tr Z^J$ to infinity (equivalently the string goes to the
center of AdS). The second operator will be at the origin, or in the
case of the Wilson loop, the loop will be a unit circle around 
the origin.

In the first case (two local operators) the transformed classical 
solution is just
\eq
z=e^{j\tau} \qqqq \phi=ij\tau
\eqx
In the second case (a local operator and the circular Wilson loop) the 
relevant solution has been found by Zarembo in \cite{CORRLOOP}:
\begin{align}
\label{e.wccorr}
x_1 &= \f{\sqj e^{j\tau}}{\cosh(\sqj \tau+\xi)} \cos \sg\nonumber\\
x_2 &= \f{\sqj e^{j\tau}}{\cosh(\sqj \tau+\xi)} \sin \sg \nonumber\\
z   &= \left( \sqj \tanh\left(\sqj \tau+\xi\right)-j \right) e^{j\tau} \nonumber\\
\phi&= ij\tau
\end{align}
where
\eq
\xi= \log \left(j+\sqj\right)
\eqx
The solution of the linear system is very simple for the first case.
It is given by
\eq
\Psi_1(w,\wb;x)=e^{-\f{j}{2(1-x)}w -\f{j}{2(1+x)}\wb} \vc{1}{0}
\quad
\Psi_2(w,\wb;x)=e^{\f{j}{2(1-x)}w +\f{j}{2(1+x)}\wb} \vc{0}{1}
\eqx
where $w=\tau+i\sg$ and $\wb=\tau-i\sg$.
A Lax matrix can be constructed immediately with $A(x)=\diag(1,-1)$.
The resulting algebraic curve is just
\eq
y^2=1
\eqx
This means that we have just two disconnected copies of the complex plane
(or the sphere).

The case of the Wilson loop correlation function is by contrast much more
complicated. The solutions of the linear system are
\begin{align}
\label{e.linearcorr}
\Psi_1 &=\, \scriptstyle \f{e^{-\f{j}{2(1-x)}w -\f{j}{2(1+x)}\wb}}{
e^{\sqj(w+\wb)}-1}
\vc{\scriptstyle \f{-i}{2\sqj} (e^{\sqj(w+\wb)}-1) \left(\red{x} \violet{-j-\sqj \left(1+ \f{2}{e^{\sqj(w+\wb)}-1}
\right)}\right)}{\scriptstyle e^{\f{1}{2}(-1+j+\sqj)w} e^{\f{1}{2}(1+j+\sqj)\wb} } \nonumber\\
\Psi_2 &=\, \scriptstyle \f{e^{\f{j}{2(1-x)}w +\f{j}{2(1+x)}\wb}}{
e^{\sqj(w+\wb)}-1}
\vc{\scriptstyle e^{\f{1}{2}(1-j+\sqj)w} e^{\f{1}{2}(-1-j+\sqj)\wb} }
{\scriptstyle \f{-i}{2\sqj} (e^{\sqj(w+\wb)}-1) \left(\red{x}\violet{-j+\sqj \left(1+ \f{2}{e^{\sqj(w+\wb)}-1}
\right)}\right)} \nonumber\\
&{}
\end{align}
Apart from being much more complicated, the above expressions are
quite surprising. Firstly, we see that once we would normalize the
vector by keeping the upper component equal to 1, the position
of the pole would move depending on the point of the worldsheet --
it would be a dynamical pole.
Yet, the pseudomomentum is trivial and characteristic of a simple
point-like string associated to a genus-0 curve and thus with
no dynamical poles. 

Let us now construct the Lax matrix and identify the corresponding
algebraic curve. It turns out that a \emph{polynomial} Lax matrix
can be constructed by taking $A(x)=(1+2jx-x^2) \cdot \diag(1,-1)$.
The expression for the resulting Lax matrix is quite involved, but yields the
relatively simple algebraic curve
\eq
y^2=(1+2jx-x^2)^2
\eqx
We see a new feature appearing -- double zeroes on the r.h.s.
So there are no true branch cuts but rather degeneracies of the
curve. We will show in the following section that these degeneracies
play a crucial role in reconstructing the solutions (\ref{e.linearcorr})
and hence also (\ref{e.wccorr}). Indeed it is worth pointing
out that the pseudomomentum $p(x)$ is not necessarily enough
to completely characterize an algebraic curve.

\section{Reconstructing the solutions from algebraic curves}
\label{s.recon}

In this section we will show how to reconstruct the solutions of the
linear system (\ref{e.linear}) from the algebraic curves identified
in the previous section.

Since we will assume that the algebraic curves came from
quite generic polynomial Lax matrices which were not associated
to any kind of monodromy, we have to rederive, following \cite{BABELON}, 
the conditions
for the essential singularity of the Baker-Akhiezer function.

The starting point is the very general fact (\cite[eq.\ (3.15)]{BABELON}),
that the flat currents $J$ and $\Jb$ can be extracted by
taking the \emph{singular} terms in the Laurent expansion
of some polynomial in $L$ with coefficients being rational
functions of $x$, i.e.
\eq
\left[ P(L(w,\wb;x),x) \right]^-_{x=1}=J(w,\wb;x)
\eqx
and similarly at $x=-1$ for $\Jb$ (taking possibly a different polynomial).

For the case of the $AdS_3$ $\sg$-model and the studied solutions,
this general rule simplifies dramatically and we always have
\eq
\label{e.ljjb}
\left[ \f{c_1}{1-x} L(w,\wb;x) \right]^-_{x=1}\!\!\!\!\!\!=J(w,\wb;x)
\quad\quad
\left[ \f{c_{-1}}{1+x} L(w,\wb;x) \right]^-_{x=-1}\!\!\!\!\!\!=\Jb(w,\wb;x)
\eqx
The constants in the above formula are arbitrary and can be changed
by a linear redefinition\footnote{In fact the constants could be also generalized
to arbitrary holomorphic and antiholomorphic functions $c_1(w)$, $c_{-1}(\wb)$
through a conformal redefinition of the worldsheet coordinate.} 
of the worldsheet coordinates $w$ and $\wb$.
They may indeed be complex (and not neccessarily complex conjugate
to each other) which then serves to pick the wanted signature of
the worldsheet, i.e.\ to use light-cone or holomorphic coordinates.

Once we have (\ref{e.ljjb}), the conditions for the essential singularity
around $x=\pm 1$ of the solutions of the linear system (\ref{e.linear})
directly follow \cite{BABELON}. 
Indeed we can rewrite $\dw \Psi+J \Psi=0$ as
\eq
\dw \Psi+ \f{c_1}{1-x}L\cdot \Psi+\mbox{regular}\cdot \Psi=0
\eqx
The second term is very simple, since we are interested in
solutions which are eigenvectors of $L$. So we may substitute
$L\cdot \Psi$ by $y(x) \Psi$ where $y(x)$ follows from the algebraic 
curve associated to $L$. 
Now around $x=1$, we can drop the last term and obtain the behaviour
\eq
\label{e.asympt}
\Psi \sim e^{- \f{c_1 y(x)}{1-x} w- \f{c_{-1} y(x)}{1+x} \wb} \cdot \mbox{regular}
\eqx
It is important to emphasize that the above asymptotics is a priori valid
\emph{only} in the neighborhood of $x=\pm 1$. 
We will see specific examples below. 

\subsection*{The null cusp Wilson loop --- reconstruction}

For the null cusp we start from the algebraic curve
\eq
y^2=1-x^2
\eqx
First let us fix the constants in (\ref{e.ljjb}). We could have
just as well left these constants arbitrary and redefined the worldsheet 
coordinates at the end of the calculation.
Here, for simplicity, we will just substitute the values corresponding
to (\ref{e.cusp1})-(\ref{e.cusp2}) from the start. We find that
\eq
c_1=\f{1+i}{2\sqrt{2}} \qqqq c_{-1}=\f{1-i}{2\sqrt{2}}
\eqx
Now the asymptotics (\ref{e.asympt}) yields
\eq
\label{e.cuspexp}
e^{-\left[ \f{1+i}{2\sqrt{2}} \sqrt{\f{1+x}{1-x}} w+
\f{1-i}{2\sqrt{2}} \sqrt{\f{1-x}{1+x}} \wb \right]}
\eqx
which exactly coincides with the overall exponential factor in 
(\ref{e.cusp1}) and (\ref{e.cusp2}). 
We will justify the form of this expression away from $x=\pm 1$
more rigorously below.

Let us now perform the reconstruction according to the procedure
of section \ref{s.finitegap}. It is first convenient to uniformize
the algebraic curve $y^2=1-x^2$ by the parametrization
\eq
y=\f{2t}{1+t^2} \qqqq x=\f{1-t^2}{1+t^2}
\eqx
In this way we get rid of all ambiguous cuts in our expressions. Passing
to the other sheet corresponds to the transformation $t \to -t$.
The points above $x=\infty$, namely $x=\infp$ and $x=\infm$
correspond to the points $t=\pm i$. The point $x=1$ corresponds to
$t=0$, while $x=-1$ corresponds to $t=\infty$.

We will first determine the normalized eigenvector of $L$ (without
using, of course, the specific form of $L(w,\wb;x)$ but only very
general properties like the diagonalizability at $x\to \infty$)
\eq
\Psi_n(w,\wb;x)=\vc{1}{\psi_n(w,\wb;x)}
\eqx
Since the genus of the algebraic curve is zero, we expect the
function $\psi_n(w,\wb;x)$ to have just a single pole.
At $x=\infty$ the Lax matrix is diagonal (can be diagonalized), so
we should have
\eq
\Psi_n(w,\wb;x=\infp)=\vc{1}{0} \quad\quad 
\Psi_n(w,\wb;x=\infm)=\vc{1}{\infty}
\eqx
This is enough to fix completely the spectral parameter dependence of
$\Psi_n(w,\wb,x)$:
\eq
\Psi_n(w,\wb;t)=\vc{1}{a(w,\wb) \f{t-i}{t+i}} \equiv
\vc{1}{-a(w,\wb) \left(x+i \sqrt{1-x^2}\right)}
\eqx
We see, that we have recovered the vector structure of 
(\ref{e.cusp1})-(\ref{e.cusp2}).

Now it remains to determine the Baker-Akhiezer function. We have already
fixed the essential singularities. Now we can justify why the expressions
in the exponent of (\ref{e.cuspexp}) are correct not only in the
neighborhood of $x=\pm 1$ but in fact for all $x$. Indeed
\eq
\sqrt{\f{1+x}{1-x}}=\f{1}{t} \qqqq \sqrt{\f{1-x}{1+x}}=t
\eqx
so these are the unique functions on the algebraic curve which have
only a single pole at $t=0$ ($x=1$) and at $t=\infty$ ($x=-1$).
Since there are no dynamical poles\footnote{I.e.\ poles whose position
depends on the worldsheet coordinates $w$ and $\wb$.} in 
$\Psi_n(w,\wb;x)$, the whole $x$ dependence of the Baker-Akhiezer
function is now fixed. So currently we have
\begin{align}
\Psi(w,\wb;t) &= f_{BA}(w,\wb;t) \cdot \Psi_n(w,\wb;t) \nonumber\\
&=
e^{-\left[ \f{1+i}{2\sqrt{2}}\f{1}{t} w+
\f{1-i}{2\sqrt{2}} t \wb \right]} b(w,\wb) \cdot 
\vc{1}{a(w,\wb) \f{t-i}{t+i}}
\end{align}
It remains to fix the functions $a(w,\wb)$ and $b(w,\wb)$.
Remarkably enough this can be done using the obvious property
that $\Psi(w,\wb,x)$ becomes $w$ and $\wb$ independent when $x\to \infty$.
This follows from the fact that then the flat currents vanish.

In our case we have to enforce this condition both at $x=\infp$ and 
at $x=\infm$.

At $x=\infp$ ($t=i$) we should impose this condition on the top component 
of $\Psi(w,\wb;t)$ and find $b(w,\wb)$:
\eq
b(w,\wb)=\f{1}{f_{BA}(w,\wb;t=i)}=e^{\f{1+i}{2\sqrt{2}} (-iw+\wb)}
\eqx
At $x=\infm$ ($t=-i$) we should concentrate on the lower component to find
$a(w,\wb)=1/b^2(w,\wb)$ which gives for the relevant product
\eq
a(w,\wb) b(w,\wb)=e^{-\f{1+i}{2\sqrt{2}} (-iw+\wb)}
\eqx
We see at this stage that we have completely recovered the solutions
of the linear system (\ref{e.cusp1})-(\ref{e.cusp2}), purely from the 
algebraic curve $y^2=1-x^2$ and some minimal assumptions on the form
of $L$ (diagonalizability at $x=\infty$ and the form (\ref{e.ljjb})).

\subsection*{The $q\bar{q}$ potential Wilson loop --- reconstruction}

In this section we will use elliptic theta functions with a square period
lattice with quasiperiods $2K=2K(\frac12),2iK'=2iK$ (for the details on notation
and properties of the doubly periodic functions, see appendix
\ref{app.elliptic}). We define the functions $x(z),y(z)$ so that $x$ has a
double pole at $iK$ and a double zero over $K$, while $y$ has zeroes over
$0,K,K+iK$ and a triple pole over $iK$. Thus, up to a multiplicative constant,
\begin{align}
x(z)&\propto\frac{\theta(z-K)\theta(z+K)}{\theta(z-iK)\theta(z+iK)}\\
y(z)&\propto\frac{\theta(z)\theta(z-K)\theta(z+K+iK)}{\theta(z-iK)\theta(z+iK)^2}.
\end{align}
They are periodic in both directions. We choose the
proportionality constant for $x$ so that $x(0)=1,x(K+iK)=-1$. Then examining
poles and zeroes on both sides of the algebraic curve equation \eqref{e.qqcurve}
\begin{equation}
\label{e.qqcurve2}
y^2=x(1-x^2)
\end{equation}
we see that they coincide, so by choosing a proportionality constant for $y$ the
above equation
can be exactly satisfied. This is thus a parameterization of this algebraic
curve, with a property that flipping the sign of $z$ corresponds to passing from
any given point to its counterpart on the other sheet of the curve (due to the
fact that $x,y$ are even and odd, respectively).

The Baker-Akhiezer prefactor has the following asymptotic structure:
\begin{align}
\label{e.qqba}
f_{BA}(w,\wb;z)=\exp\left\{\frac{-i}{2\sqrt2}\left(\frac{y}{1-x}w+\frac{-y}{1+x}\wb\right)\right\}\times\mbox{regular}
\end{align}
where again the constants were specifically chosen but in principle could have
been redefined at the very end. However, we will choose to work with
$\sigma,\tau$ instead of $w=\sg+i\tau,\wb=\sg-i\tau$, motivated by the fact 
that the original
solution for this case \eqref{e.qqpsi1}-\eqref{e.qqpsi2} was more conveniently
expressed in terms of these. 
We have to ensure that the Baker-Akhiezer prefactor does not have any
essential singularities other than $x=\pm1$ (i.e.\ $z=0$ or $z=K+iK$).
We see, however, that the exponent in (\ref{e.qqba}) has the following poles:
\begin{align}
\frac{y}{1-x}&=i\frac{\sqrt2}{z}-i\frac{\sqrt2}{z-iK}+\mbox{regular},\\
\frac{-y}{1+x}&=i\frac{\sqrt2}{z-K-iK}-i\frac{\sqrt2}{z-iK}+\mbox{regular},
\end{align}
so in their difference (multiplied by $\tau$ in the exponent) the second terms
will cancel out. Hence the $\tau$-dependent part becomes
\eq
\exp \left\{ \f{-i}{2\sqrt{2}} \left( \f{y}{1-x}+\f{y}{1+x} \right) i\tau \right\}
=\exp \left\{ \f{x\tau}{\sqrt{2}y} \right\}
\eqx
However in the coefficient of $\sg$, the pole at $z=i K$ corresponds to $x=\infty$
and as such is
forbidden in the Baker-Akhiezer function properties outlined 
in section~\ref{s.finitegap}. 
The function multiplying $\sg$ in the exponent will have to have a pole at 
$z=0$ and $z=K+iK$ with the prescribed residues. Such a function can
be explicitly constructed as
\begin{equation}
G(z)=-\smallhalf(\dlogth(z)+\dlogth(z-K-iK)).
\end{equation}
where $\dlogth(z)$ is the logarithmic derivative of $\th(z)$ 
(see appendix~\ref{app.elliptic}).

Now $f_{BA}$ is no longer periodic in the imaginary direction as the residues
do not sum up to zero, and to remedy this we supply it with another factor
\begin{align}
\frac{\theta(z-\gamma(\sigma,\tau))}{\theta(z-\gamma(0,0))}.
\label{e.qqdynpole}
\end{align}
$\gamma(\sg,\tau)$ denotes the position of the dynamical pole (as $f_{BA}$
has to vanish there). Demanding the double periodicity of $f_{BA}$ fixes
the position of the dynamical pole to $\gamma=-i\sigma$ (with $\gamma(0,0)=0$).

The ansatz for the solution of (\ref{e.linear}) becomes
\begin{equation}
\Psi=A(\sg,\tau) \cdot \exp\left(\frac{x\tau}{\sqrt2y}+iG(z)\sigma\right)
\frac{\theta(z+i\sigma)}{\theta(z)}\vc{1}{\psi(\sigma,\tau;z)}
\end{equation}
where $A(\sg,\tau)$ is the $\sigma,\tau$-dependent regular part of $f_{BA}$. 
The function
$\psi$ should have its poles at $x=\infty$ and at $z=\gamma$ and should be a well
defined function on the elliptic curve (\ref{e.qqcurve2}).
It can be constructed in the following form
\begin{align}
\psi(\sigma,\tau;z)=r_0(\sg,\tau)+r_1(\sg,\tau)(\dlogth(z-iK)-
\dlogth(z+i\sigma))
\end{align}
where we have chosen the residues at both poles to cancel to ensure periodicity.

We now determine the unknown functions according to the requirement that $\Psi$
be constant at $x=\infty$ and, due to the fact that $x(z)$ has a double pole at
$z=iK$, we have to require that $\Psi$ is constant at the two leading orders in
the expansion around $z=iK$. Using the expansions
\begin{align}
f_{BA}&=A(\sg,\tau)(f_0(\sg,\tau)+(z-iK)f_1(\sg,\tau)+\ldots),\\
\psi&=\frac{\psi_{-1}(\sg,\tau)}{z-iK}+\psi_0(\sg,\tau)+\ldots,
\end{align}
we can write
\begin{align}
\Psi=A(\sg,\tau)\left(\begin{array}{c@{\ +\ }c@{\ +\ldots}}0&f_0(\sg,\tau)
\\\frac{f_0(\sg,\tau)\psi_{-1}(\sg,\tau)}{z-iK}&\psi_{-1}(\sg,\tau)f_1(\sg,\tau)
+f_0(\sg,\tau)\psi_0(\sg,\tau)\end{array}\right)
\end{align}
and demand that all the above coefficients be constant at $z=iK$. 
We obtain the following solution:
\begin{align}
A(\sg,\tau)&=\frac{C_1}{f_0(\sg,\tau)} \qqqq 
r_1(\sg,\tau)=\psi_{-1}(\sg,\tau)=\frac{C_2}{C_1}\\
\psi_0(\sg,\tau)&=\frac{C_3-A(\sg,\tau)\psi_{-1}(\sg,\tau)f_1(\sg,\tau)}{A(\sg,\tau)f_0(\sg,\tau)}=\frac{C_3}{C_1}-
\frac{f_1(\sg,\tau)}{f_0(\sg,\tau)}\cdot\frac{C_2}{C_1}
\end{align}
and $r_0$ (contained in $\psi_0$) is then
\begin{align}
r_0(\sg,\tau)=\frac{C_3}{C_1}-\frac{f_1(\sg,\tau)}{f_0(\sg,\tau)}\cdot\frac{C_2}{C_1}-
\frac{C_2}{C_1}\left(\frac{\theta''(0)}{2\theta'(0)}-
\dlogth(iK+i\sigma)\right).
\end{align}

This is of course at first sight very different from
\eqref{e.qqpsi1}-\eqref{e.qqpsi2}, but some agreement is to be expected, firstly
due to the fact that the Jacobi and theta functions are related (albeit very
intricately). Secondly, we might notice that regardless of the value of $z$,
$\Psi$ ceases to be well defined at $\sigma=\pm K$, due to the factor
$\theta(iK+i\sigma)$ that is present in $f_0$, a denominator of $A$. This means
that the domain of this solution is $\sigma\in(-K,K)$, precisely the same as for
the original solution. Note that this actually follows from a specific choice of
$\gamma(0,0)$, as alternative values would shift the domain
or lead generically to complex solutions.

Finally, for a specific choice of the constants $C_{1,2,3}$ we get exact
agreement. If we choose $C_1=\sqrt{1-x},C_2/C_1=i\sqrt2,C_3=0$, then the result
is numerically equal to \eqref{e.qqpsi1} (with replacements
$x\to x(z),\sqrt x\sqrt{1-x^2}\to y(z)$) up to hundreds of decimal digits 
for all $\sigma,\tau,z$. 


\subsection*{The $\cor{W(C)\tr Z^J}$ correlation function --- reconstruction}

Here we start from the degenerate curve
\eq
\label{e.degen}
y^2=(1+2jx-x^2)^2
\eqx
In this case it is not completely obvious what conditions to impose
on the analytic structure of the solutions of the linear system.
A point of view which we will adopt here will be to consider the
curve (\ref{e.degen}) as a degenerate limit of a curve with two
very small cuts. Thus we may treat it as a degenerate limit of
an elliptic curve. Fortunately, we do not need to perform the elliptic
construction first and only at the end take the limit --- we may directly
work with the degenerate curve, drawing from the genus-1 case only
some very general analyticity properties. However, for this degenerate
curve, we cannot rule out the existence of some other different 
constructions.

Firstly, the two sheets of (\ref{e.degen}) are completely distinct
and there is no analytical continuation between them. Hence we may,
and should, consider two separate vector functions for the two
independent linear solutions of (\ref{e.linear}). Secondly, as we
may expect the curve to come as a degeneration of an elliptic curve,
we expect to have one kinematical pole at $x=\infty$, and one
dynamical pole (depending on $w$ and $\wb$). We have to distribute
those two poles between the two branches. Thirdly, at the points
of degeneration $1+2jx-x^2=0$, we will require the two solutions
to coincide.

Let us start from the essential singularities at $x=\pm 1$.
In this case we find the constants to be
\eq
c_1=-\f{1}{4} \qqqq c_{-1}=\f{1}{4}
\eqx  
which gives the behaviour
\eq
\Psi \sim e^{\f{1}{4}\f{ y(x)}{1-x} w- \f{1}{4} \f{ y(x)}{1+x} \wb} 
\cdot \mbox{regular}
\eqx
However care must be taken here, since $y(x)=\pm (1+2jx-x^2)$.
We cannot substitute this full expression into the exponent since
this would generate an unwanted essential singularity at $x=\infty$.
Hence it is simplest to just substitute $y(1)$ in the first term and
$y(-1)$ in the second term.\footnote{A possible piece proportional
to $(x-1)$ or $(x+1)$ would be automatically cancelled later in the
calculation.} We get therefore
\eq
\Psi \sim e^{\pm \left(\f{1}{2} \f{j}{1-x} w+\f{1}{2} \f{j}{1+x} \wb\right)}
\eqx

Now we have to distribute the poles among the two solutions. We will put the
pole at $x=\infty$ in the first solution and the dynamical pole in the second.
This choice leads to the following ansatz:
\begin{align}
\Psi_1(w,\wb;x) &= e^{\f{1}{2} \f{j}{1-x} w+\f{1}{2} \f{j}{1+x} \wb}
c_1(w,\wb) \vc{1}{b_1(w,\wb) (x-a_1(w,\wb))} \\
\Psi_2(w,\wb;x) &= e^{-\f{1}{2} \f{j}{1-x} w-\f{1}{2} \f{j}{1+x} \wb}
c_2(w,\wb) (x-a_2(w,\wb))
\vc{1}{\f{b_2(w,\wb)}{x-a_2(w,\wb)}}
\end{align}
Now we impose the condition that at $x\to \infty$, the solution becomes
$w$, $\wb$ independent. This gives the relations $b_1(w,\wb)=1/c_1(w,\wb)$
and $c_2(w,\wb)=1$. So at this stage our ansatz takes the form
\begin{align}
\Psi_1(w,\wb;x) &= e^{\f{1}{2} \f{j}{1-x} w+\f{1}{2} \f{j}{1+x} \wb}
\vc{c_1(w,\wb)}{x-a_1(w,\wb)} \\
\Psi_2(w,\wb;x) &= e^{-\f{1}{2} \f{j}{1-x} w-\f{1}{2} \f{j}{1+x} \wb}
\vc{x-a_2(w,\wb)}{b_2(w,\wb)}
\end{align}
Finally, since we expect that the two different functions should come
from the same function on the (almost degenerate) elliptic curve,
we require that at the two points of degeneration
\eq
x=j \pm \sqrt{1+j^2}
\eqx
we have
\begin{align}
\Psi_1\left(w,\wb;j+\sqrt{1+j^2}\right) &= \Psi_2\left(w,\wb;j+\sqrt{1+j^2}\right) \\
\Psi_1\left(w,\wb;j-\sqrt{1+j^2}\right) &= \Psi_2\left(w,\wb;j-\sqrt{1+j^2}\right)
\end{align}
This gives a set of four linear equations for the four unknown
functions $a_1(w,\wb)$, $a_2(w,\wb)$, $b_2(w,\wb)$ and $c_1(w,\wb)$.
The solution is
\begin{align}
a_1(w,\wb) &= j-\sqrt{1+j^2}\left(1-\frac2{1-e^{(w+\wb)\sqrt{1+j^2}}}\right) \\
a_2(w,\wb) &= j+\sqrt{1+j^2}\left(1-\frac2{1-e^{(w+\wb)\sqrt{1+j^2}}}\right) \\
b_2(w,\wb) &= -\frac{2\sqrt{1+j^2}e^{-\frac w2(1-j-\sqrt{1+j^2})+\frac\wb 2(1+j+\sqrt{1+j^2})}}{1-e^{(w+\wb)\sqrt{1+j^2}}} \\
c_1(w,\wb) &=  \frac{2\sqrt{1+j^2}e^{ \frac w2(1-j+\sqrt{1+j^2})-\frac\wb 2(1+j-\sqrt{1+j^2})}}{1-e^{(w+\wb)\sqrt{1+j^2}}} 
\end{align}
and coincides with the quite intricate expressions (\ref{e.linearcorr}) 
for the solution of the linear system.

\section{Applications and conclusions}
\label{s.appl}

The aim of this paper was to show that the classical 
algebraic curve (finite-gap) classification of spinning string
solutions in $AdS_5 \times S^5$ can be significantly 
expanded to encompass other more general classes of solutions,
namely Wilson loops and, possibly, correlation functions.

The first case is perhaps not surprising from the point of
view of the classical literature on minimal surfaces and
integrable models \cite{Bobenko}, although the focus there has been
always rather different and the kind of minimal surfaces
relevant for computing Wilson loop expectation values within
the AdS/CFT correspondence did not appear. However, it
definitely points at a new direction in the context of
the spinning string classification, as all these Wilson
loops have no noncontractible loops, hence no monodromy
and no pseudomomentum $p(x)$, whose analytic properties
were the starting point for the spinning string
classification \cite{KMMZ,KZ,BKSZ,ALGREVIEW}. 

In this paper we showed that for certain classical
Wilson loop minimal surfaces in $AdS_3$, namely the one associated
with a null cusp and the infinite rectangular Wilson loop 
responsible for the $q\bar{q}$ potential, there exists
an underlying algebraic curve description. We can associate a definite
algebraic curve with each of these solutions and conversely,
starting just from that algebraic curve, we can reconstruct
the explicit target-space form of the classical solution.

These results have, on the one hand, a purely practical 
application of suggesting
new methods of constructing minimal surfaces in an Anti-de-Sitter
spacetime starting from some higher genus algebraic curves.
In this respect, it would be very interesting to understand the
precise relation (or even perhaps equivalence) with the very 
interesting constructions of \cite{Kruczenski}.
On the other hand, the main motivation for us was more theoretical,
as the finite-gap constructions of spinning strings had
a lot in common with Bethe equations and the comparison
with analogous constructions for weak coupling spin chains
played a very important role in the development of
integrability.

It would be very interesting to understand if there is
a similar underlying Bethe ansatz interpretation of the
Wilson loops with an algebraic curve description, and in
particular understand the relation with the very recent 
works \cite{CHMSqq,Drukkerqq}. 

On a less speculative level, from the perspective of algebraic
curves we may understand quite easily
the possible limit-like relations between various string solutions.
Of particular interest is the very close relation of the null
cusp solution with the large spin limit of the GKP folded string \cite{KRUCZGKP}.
This relation is especially important, as the GKP string
is a closed string solution which is describable \emph{at all
couplings} by the all-loop Bethe ansatz\footnote{The Bethe ansatz
description is valid in the large spin 
limit. For generic spins, the description would be in terms of
TBA/NLIE equations.} \cite{ES,BES}. Let us see how this
relation arises from the point of view of the identified
algebraic curves.

The null cusp is described by the curve $y^2=(x^2-1)$. One
can make a deformation of the above curve by adding two additional
branch points and taking them to infinity. This suggests
to consider the curve
\eq
y^2=(x^2-1)(x^2-a^2)
\eqx
in the $a \to \infty$ limit. As we show in appendix \ref{s.gkp},
this curve is indeed the algebraic curve underlying the GKP folded
string.

A natural very interesting question is whether a similar relation
exists for the Wilson loop describing the $q\bar{q}$ potential, i.e.
whether there exists a (closed string) solution which would approximate
in some form the $q\bar{q}$ minimal surface. To this end
we should deform the algebraic curve $y^2=x(x^2-1)$.
A natural choice would be to use the curve
\eq
y^2=(x^2-1)(x-a)(x+1/a)
\eqx
and take the limit $a \to \infty$. In appendix \ref{s.genqq},
we identify the corresponding string solution. Unfortunately it turns
out to be also a Wilson loop minimal surface -- namely the generalized 
Wilson loop of two parallel lines on the boundary of global $AdS_3$
with an angular separation. This configuration indeed has been 
proposed in \cite{FD} as a generalization of the ordinary $q\bar{q}$
potential and used very recently in \cite{CHMSqq,Drukkerqq}.
Unfortunately we do not find a closed string counterpart. However,
we cannot rule out that some complexified version with fine-tuned
parameters (or some genus-2 degeneration) could exist.

The second line of generalization of the classical finite-gap
constructions is the case of correlation functions with a local
operator. For these classical solutions, the monodromy around the
puncture where the local operator would be inserted should be,
by definition, \emph{identical} to the monodromy of the
corresponding spinning string.  Hence the ordinary classical
algebraic curve which is constructed out of the pseudomomentum
would be identical to the one for the spinning string.

Yet clearly, there is a multitude
of nonvanishing correlation functions in which even the simplest
operator like the BMN vacuum $\tr Z^J$ could participate.
This suggests that the space of solutions with given
pseudomomentum around a puncture should be extremely vast.
This is in a naive contradiction with the folklore that
the space of classical solutions of a genus-$g$ algebraic
curve is finite dimensional. 

We address this problem by examining a simple example
of a correlation function of the circular Wilson loop
with the BMN operator $\tr Z^J$. We find that even though
the pseudomomentum is the same, the algebraic curve
constructed from a polynomial Lax matrix is
singular and can be treated as a degeneration
of an elliptic curve. The singularities play a key role
in the reconstruction of the Wilson loop correlator from
the algebraic curve. Intuitively, the solution may be
understood as a soliton (degenerate cuts) on top of
a finite-gap spinning string. It would be interesting to
explore these types of constructions for other local
operators/spinning string solutions.

Clearly, in the case of correlation functions this result
is just scratching the surface. For 3-point correlation functions,
we expect the classical string solutions to be
\emph{simultaneously} describable by three distinct
algebraic curves, even of different genera. Currently,
we do not possess even a single example (even in some
simplified integrable model) with such characteristics.
It would be very interesting to understand the structure
of such solutions from the algebraic curve perspective.

\bigskip

{\bf Acknowledgements.}
This work is supported by the International PhD Projects Programme of the
Foundation for Polish Science within the European Regional Development 
Fund of the European Union, agreement no.\ MPD/2009/6,
as well as by Polish science funds as a research 
project N N202 105136 (2009-2012).
RJ was supported by the Institute for Advanced Studies, Jerusalem 
within the Research Group {\it Integrability and Gauge/String Theory.}
PLG thanks the
Laboratory of Theoretical Physics of École Normale Superieure in Paris for
hospitality during the period when a part of this work has been performed.


\appendix

\section{Useful elliptic functions}
\label{app.elliptic}

In this appendix we will review certain basic properties of both Jacobi elliptic
functions and theta functions. We will largely limit the scope to the properties
essential to our calculations; for a more comprehensive discussion, including
different notations encountered in the literature, refer eg.\ to the relevant
chapters of \cite{STEGUN} or \cite{NIST}.

The {\bf Jacobi elliptic functions} are defined (in one of many equivalent ways)
as follows: if
\begin{equation}
u=F(\varphi|m)=\int_0^\varphi\frac{{\rm d}\theta}{\sqrt{1-m\sin^2\theta}}
\end{equation}
where $F(\varphi|m)$ is the incomplete elliptic integral of the first kind, then
\begin{align}
\am u&=\varphi & \sn u&=\sin\varphi\\
\cn u&=\cos\varphi & \dn u&=\sqrt{1-m\sin^2\varphi}
\end{align}
where the first function is called the (Jacobi) amplitude. 
The number $m$ is a second, usually suppressed,
argument to all of the functions, called parameter (as opposed to an alternative
notation which uses its square root, called modulus, instead).

Among the elementary properties of the Jacobi elliptic functions are the relations
between their square roots (directly following from the above definitions and
the trigonometric unity):
\begin{equation}
\sn^2u+\cn^2u=m\sn^2u+\dn^2u=1.
\end{equation}
Essential to our derivations are their derivatives (with respect to the
non-suppressed argument) as well:
\begin{equation}
\sn'u=\cn u\dn u \qquad \cn'u=-\sn u\dn u \qquad \dn'u=-m\sn u\dn u.
\end{equation}
We have also used the formulas (note that they apply to the case $m=\frac12$ only):
\begin{equation}
[E(\am u|\smallhalf)]'=\smallhalf(\cn^2u+1)\qquad
[\Pi({\textstyle\frac{x}{x-1}};\am u|\smallhalf)]'=\displaystyle\frac{1-x}{1-x\cn^2 u}.
\end{equation}

The {\bf theta functions} are a collection of four special functions defined via
their Fourier expansions; here we will use only one of them, namely
\begin{equation}
\theta_3(z|\tau)=1+2\sum_{n=1}^\infty q^{n^2}\cos(2nz),
\end{equation}
where $\tau$ ($\Im\tau>0$) is the lattice parameter, once it is chosen, 
it is usually suppressed. $q=e^{i\pi\tau}$ is the nome.
The quasi-periodicity in case of $\theta_3$ is expressed as
\begin{equation}
\theta_3(z+(m+n\tau)\pi)=q^{-n^2}e^{-2inz}\theta_3(z)
\end{equation}
for integer $n,m$ (and thus $\pi$ is an actual, not only quasi, period).

This property allows one to very easily construct meromorphic 
functions on the elliptic 
curve (i.e.\ doubly periodic functions on the complex plane).
Indeed, functions of the following types
\begin{equation}
\prod_{i=1}^n\frac{\theta_3(z-a_i)}{\theta_3(z-b_i)} \qqqq \sum_{i=1}^n R_i\partial_z\ln\theta_3(z-b_i)\label{e.dblper}
\end{equation}
are actually doubly periodic (not only quasi) under the following conditions:
$\sum a_i-\sum b_i=k\pi$ for integer $k$, and $\sum R_i=0$, respectively. 
The first form is very convenient to use if we have information on the location
of zeroes and poles of the elliptic function that we want to construct, while
the second form is convenient if the function has only single poles with
prescribed residues.

Since $\theta_3$ has a zero at $z=\pi (1+\tau)/2$, it is convenient to 
shift the argument in order to have a function which vanishes at $z=0$.
We denote such a function by $\th(z)$ and its logarithmic derivative
by $\phi(z)=\partial\ln\theta(z)$. In addition it is sometimes convenient
to also rescale the argument. We use explicitly
\begin{align}
\theta(z)&=\theta_3\left(\frac{\pi z}{k}-\frac{1+\tau}{2}\pi\bigg|\tau\right)\\
\theta(z+k(m+n\tau))&=e^{-i\pi n(\tau n-1-\tau+2z/k)}\theta(z)\\
\phi(z)&=\frac{\pi}{k\theta(z)}\theta'_3\left(\frac{\pi z}{k}-\frac{1+\tau}{2}\pi\bigg|\tau\right)
\end{align}
with $k=2K(\frac12),\tau=i$ in the case of $q\bar q$ reconstruction and $k=2\om$ 
and $\tau=\om'/\om$ in the following appendix. 
Note that the expressions of the form \eqref{e.dblper},
but with $\theta$ instead of $\theta_3$, have poles precisely at all $b_i$'s
(and zeroes at $a_i$'s in the former case).

\section{Other elliptic reconstructions}

In this appendix, we will argue how one can reconstruct
the GKP folded string from the curve $y^2=(x^2-1)(x^2-a^2)$
and the generalized $q\bar{q}$ potential from 
$y^2=(x^2-1)(x-a)(x+1/a)$ giving justification
to the statements made in section \ref{s.appl}. We start from 
giving some very general formulas which we then specialize 
to the two curves of interest.

\subsection{Generalities}

Let us briefly review the generic features of reconstructing
the classical solution from a general elliptic curve, with
the proviso that the Lax matrix is diagonal at $x=\infty$
(so the situation is simpler than for the case of $q\bar{q}$
potential discussed in the main text), and $x=\infty$ is
not a branch point of the algebraic curve.

We will denote the (spectral) coordinate on the elliptic
curve by $u$. We will always assume that the passage to
the other sheet occurs through the transformation $u \to -u$,
i.e.
\eq
x(-u)=x(u) \qqqq y(-u)=-y(u)
\eqx
The branch points will then be located at the half-periods
\eq
u=0,\om,\om',\om+\om' \quad \text{or} \quad
u=0,\f{1}{2},\f{\tau}{2},\f{1+\tau}{2}
\eqx
For solutions completely contained in $AdS_3$, two of these branch
points will correspond to $x=+1,-1$. We will denote these positions
by $u=\plone,\mnone$. The other points of relevance on the elliptic curve
are the images of $x=\infty$: $u=\infp$ and $u=\infm \equiv -\infp$;
and the images of $x=0$: $u=\zerp$ and $u=\zerm \equiv -\zerp$.

From the discussion in the main text we know that the lower component
of the normalized eigenvector $\Psi_n(w,\wb;u)$ should have a zero at
$u=\infp$, a pole at $u=\infm$ and a further single dynamical pole.
Consequently it can be written as
\eq
\Psi_n(w,\wb;u) = \vc{1}{b(w,\wb) \f{\th(u-\infp)}{\th(u-\infm)} \cdot
\f{\th(u+\infp-\infm-\gm)}{\th(u-\gm)}}
\eqx
where $\gm \equiv \gm(w,\wb)$ is the position of the single dynamical pole.
The Baker-Akhiezer function is again immediate to write:
\eq
f_{BA}(w,\wb;u)=a(w,\wb) \cdot e^{ \dlogth(u-\plone) w + \dlogth(u-\mnone) \wb }
\cdot \f{\th(u-\gm)}{\th(u-\gm_0)}
\eqx
where $\gm_0$ is some reference point. The requirement that $f_{BA}$ is
doubly periodic in $u$ allows us to determine the position of the dynamical
pole, as in \eqref{e.qqdynpole}. We obtain
\eq
\gm(w,\wb)=w+\wb+\gm_0
\eqx
$a(w,\wb)$ and $b(w,\wb)$ may be easily reconstructed from the behaviour
at $u=\infp$ and $u=\infm$. The result is
\eq
\Psi(w,\wb;u)=\vc{
\f{e^{ \dlogth(u-\plone) w + \dlogth(u-\mnone) \wb}}{
e^{ \dlogth(\infp-\plone) w + \dlogth(\infp-\mnone) \wb}}
\cdot
\f{\th(\infp-\gm_0) \th(u-\gm)}{\th(u-\gm_0) \th(\infp-\gm)}}{
b\f{e^{ \dlogth(u-\plone) w + \dlogth(u-\mnone) \wb}}{
e^{ \dlogth(\infm-\plone) w + \dlogth(\infm-\mnone) \wb}} 
\cdot
\f{\th(\infm-\gm_0) \th(u-\infp) \th(u+\infp-\infm-\gm)}{
\th(u-\gm_0) \th(u-\infm) \th(\infp-\gm)}}
\eqx
where $b$ is now a constant. Denoting for simplicity the two
components by
\eq
\Psi(w,\wb;u)=\vc{ \UP(u) }{b \cdot \DN(u)}
\eqx
we can put $\Psihat$ to be equal to\footnote{This is not the most general
expression but will suffice for the examples in the appendices.}
\eq
\Psihat= \arr{A_1\cdot \UP(\zerp)}{A_2\cdot \UP(\zerm)}{
A_1\cdot b\cdot \DN(\zerp)}{A_2\cdot b\cdot \DN(\zerm)}
\eqx
with $A_{1,2}$ and $b$ arbitrary constants.

We can now obtain explicit expressions for the solution in
global $AdS_3$ spacetime by using (\ref{e.gpsihat}) and the
global $AdS_3$ formula in (\ref{e.group}). We get
\begin{align}
\label{e.expt}
e^{2i t} &= \f{A_2 b}{A_1} \cdot  \f{\DN(\zerm)}{\UP(\zerp)} \\
\label{e.exppsi}
e^{2i \psi} &= \f{A_2}{b A_1} \cdot  \f{\UP(\zerm)}{\DN(\zerp)} \\
\label{e.coshrho}
\cosh^2 \rho &= \f{A_1 A_2 b \cdot \DN(\zerm) \UP(\zerp)}{\det \Psihat}
\end{align}
In the last equation $\det \Psihat$ is just a pure number.

With these expressions in hand, we will now indicate
how the well known solutions -- the GKP string and the 
generalized $q\bar{q}$ potential arise from their algebraic curves.
Of course, these solutions are much simpler to obtain directly.
For us the main motivation for doing this calculation is to
make a clear link with algebraic curves.
However, once we would want to obtain solutions of the linear
system for the GKP string, we believe that this route is the best
(as we failed to directly solve (\ref{e.linear}) for the GKP
folded string).

\subsection{$y^2=(x^2-1)(x^2-a^2)$ --- the GKP folded string}
\label{s.gkp}

We can uniformize the algebraic curve $y^2=(x^2-1)(x^2-a^2)$ either
using $\th$ functions, as in the main text, or using Weierstrass
$\wp$ functions after bringing the curve to the standard Weierstrass
form. For completeness we will give explicit formulas here.
For the present case we find
\begin{align}
x(u) &= \f{-1+5a^2+6(1-a^2)X(u)}{-5+a^2-6(1-a^2) X(u)} \\
y(u) &= \sqrt{\f{a^2-1}{2}} \cdot (1+x(u)^2) \cdot Y(u)
\end{align}
with $X(u)=\wp(u;\{g_2,g_3\})$, $Y(u)=\wp'(u;\{g_2,g_3\})$, where
\eq
g_2=\f{1 + 14 a^2 + a^4}{3(-1 + a^2)^2} 
\qqqq 
g_3=\f{1 - 33 a^2 - 33 a^4 + a^6}{27 (-1 + a^2)^3}
\eqx
We find then that $\mnone=0$, $\plone=\om$, while $\infp=\om'-\om/2$
and $\zerp=-\om/2$. In order to identify the solution with the folded
GKP string it is really enough to just check that $t=const\cdot \tau$, 
$\psi=const' \cdot \tau$ and $\rho=\rho(\sg)$.

Let us first consider $t$ and identify the worldsheet dependence
following from (\ref{e.expt}). Apart from the exponent we have
the following combination of $\th$ functions:
\eq
\f{\th(\zerm+\infp-\infm-\gm)}{\th(\zerp-\gm)}
\eqx
Using the explicit locations of these points given above,
we find that $\zerm+\infp-\infm=\zerp+2\om'$ and hence the
two $\th$ functions cancel leaving just an additional exponent.
Collecting the exponents together we find
\eq
e^{2i t} = \widetilde{const} \cdot e^{const \cdot(\wb-w)}
\eqx
We can get rid of $\widetilde{const}$ through a judicious
choice of the constants $A_{1,2}$ and $b$. This establishes
that $t=const\cdot \tau$.

For (\ref{e.exppsi}) we get similarly
\eq
\f{\th(\zerm-\gm)}{\th(\zerp+\infp-\infm-\gm)}
\eqx
Again we find that $\zerp+\infp-\infm=\zerm-2\om+2\om'$
and the same reasoning applies. Consequently we find that
\eq
e^{2i\psi} = \widetilde{const'} \cdot e^{const' \cdot(\wb-w)}
\eqx
showing that indeed $\psi=const'\cdot \tau$.

Finally, for $\cosh^2 \rho$ we find nontrivial dependence
on $w+\wb$ coming both from the $\th$ functions and from 
the exponential factor. So we get $\rho=\rho(\sg)$.
This is enough to identify the solution with the GKP folded
string.

\subsection{$y^2=(x^2-1)(x-a)(x+1/a)$ --- the generalized\\ 
$q\bar{q}$ potential}
\label{s.genqq}

The algebraic curve can be uniformized similarly as before.
We get 
\begin{align}
x(u) &=\f{3+4 a-3 a^2-6 (-1+a^2)X(u)}{3-4a-3 a^2+6(-1+a^2)X(u)} \\
y(u) &= -\sqrt{\f{a^2-1}{2a}} \cdot (1+x(u))^2 \cdot Y(u)
\end{align}
where $X(u)=\wp(u;\{g_2,g_3\})$, $Y(u)=\wp'(u;\{g_2,g_3\})$ but now with
\eq
g_2= \f{(3+a^2)(1+3a^2)}{3(a^2-1)^2}
\qqqq 
g_3= \f{-2a (9a^4+14a^2+9)}{27 (a^2-1)^3}
\eqx
We find that $\mnone=0$, $\plone=\om+\om'$, however in the present case 
$\infp$ is not given
in any simple form in terms of the half-periods. However due
to the symmetry $x \to -1/x$ of the algebraic curve, which is
realized as $u \to \om+\om' \pm u$, we can express $\zerp$ also
in terms of $\infp$ ($\infm \equiv -\infp$ and $\zerm \equiv -\zerp$
follow immediately):
\eq
\zerp=\om+\om'+\infp
\eqx 
Let us now repeat the analysis done for the GKP string. For $e^{2i t}$,
our conclusion is unchanged since again
\eq
\zerm+\infp-\infm=\infp-\om-\om' \equiv \zerp-2\om-2\om'
\eqx
so the $\th$ functions cancel. The exponents again lead 
to $t=const (\wb-w) \propto \tau$.

The situation for $e^{2i\psi}$ is, however, more subtle.
We find
\eq
e^{2i\psi} = \widetilde{const'} \cdot
\f{\th(\zerm-\gm)}{\th(\zerp+\infp-\infm-\gm)} \cdot
e^{const' (w+\wb)}
\eqx
Firstly this is now a function of $w+\wb$ instead of $\wb-w$ as
for the GKP string. Secondly, the $\th$ functions no longer cancel
and the dependence on $\sg$ is quite nontrivial.
Thirdly, we find that the requirement of a \emph{real} solution,
which corresponds here to requiring that $|e^{2i\psi}|=1$ for
some choice of constants severly restricts the choices of $\gm_0$
and the real form of the worldsheet coordinates (recall that
$\gm=w+\wb+\gm_0$). Some (nonexhaustive) numerical experimentation
leads to the choices that i) $w+\wb=2i\sg$ and ii) $\gm_0=\infp$
or $\gm_0=\zerp$. In these cases we get a real $\psi=\psi(\sg)$.

Let us now proceed to the formula for $\cosh^2 \rho$.
Here we find
\eq
\cosh^2 \rho = const \cdot \f{\th^2(\zerm+\infp-\infm-\gm)}{\th^2(\zerp-\gm)}
\cdot e^{-\f{\pi}{\om} (w+\wb)}
\eqx
Firstly we see that again this is a function of $\sg$ alone.
For the case of $\gm_0=\infp$ the expression turns out to be real
and positive. However the $\th$ function in the denominator will
have zeroes, which shows that the solution has $\rho\to \infty$ there,
which means that it reaches the boundary and hence represents
a Wilson loop. Moreover, one can check that the boundary values
of $\psi$ at the two edges differ. So the solution with $\gm_0=\infp$
corresponds exactly to a Wilson loop in global $AdS_3$, where the 
boundary lines have some angular separation. This is exactly
the case of the generalized $q\bar{q}$ Wilson loop considered in
\cite{FD,Drukkerqq,CHMSqq} which approximates the ordinary $q\bar{q}$ 
potential Wilson loop.
This identification is consistent with viewing the approximation 
on the level of algebraic curves as discussed in section \ref{s.appl}.

The second choice $\gm_0=\zerp$, which does not lead to singularities 
in the $\th$ functions, unfortunately leads to $\cosh^2 \rho<0$.
Moreover, even complexified, this solution is not periodic so cannot
be used as a counterpart of the GKP folded string for the $q\bar{q}$
potential.

\newcommand \arxiv [2][] {\href{http://arxiv.org/abs/#2}{\tt arXiv:\hspace{0pt}#2\ifthenelse{\equal{#1}{}}{}{ [#1]}}}


\begin{thebibliography}{99}

\bibitem{INTREVIEW}
  N.~Beisert, C.~Ahn, L.~F.~Alday, Z.~Bajnok, J.~M.~Drummond, L.~Freyhult, N.~Gromov, R.~A.~Janik {\it et al.},
  ``Review of AdS/CFT Integrability: An Overview,''
  Lett.\ Math.\ Phys.\  {\bf 99} (2012) 3,
  \arxiv[hep-th]{1012.3982}

\bibitem{TBA1}
  N.~Gromov, V.~Kazakov, A.~Kozak, P.~Vieira,
  ``Exact Spectrum of Anomalous Dimensions of Planar $\mathcal{N} = 4$ Supersymmetric Yang-Mills Theory: TBA and excited states,''
  Lett.\ Math.\ Phys.\  {\bf 91} (2010) 265,
  \arxiv[hep-th]{0902.4458}


\bibitem{TBA2}
  D.~Bombardelli, D.~Fioravanti, R.~Tateo,
  ``Thermodynamic Bethe Ansatz for planar AdS/CFT: A Proposal,''
  J.\ Phys.\ A: Math.\ Theor.\ {\bf 42} (2009) 375401,
  \arxiv[hep-th]{0902.3930}


\bibitem{TBA3}
  G.~Arutyunov, S.~Frolov,
  ``Thermodynamic Bethe Ansatz for the $AdS_5\times S^5$ Mirror Model,''
  JHEP {\bf 0905} (2009) 068,
  \arxiv[hep-th]{0903.0141}

\bibitem{NLIE1}
  N.~Gromov, V.~Kazakov, S.~Leurent, D.~Volin,
  ``Solving the AdS/CFT Y-system,''
  \arxiv[hep-th]{1110.0562}

\bibitem{NLIE2}
  J.~Balog, Á.~Hegedűs,
  ``Hybrid-NLIE for the AdS/CFT spectral problem,''
  \arxiv[hep-th]{1202.3244}

\bibitem{KMMZ} V.~Kazakov, A.~Marshakov, J.~Minahan, K.~Zarembo,
  ``Classical\hspace{0pt}/\hspace{0pt}quantum integrability in AdS/CFT,'' JHEP {\bf0405} (2004) 024,
  \arxiv{hep-th/0402207}

\bibitem{KZ} V.~Kazakov, K.~Zarembo, ``Classical/quantum integrability in
  non-compact sector of AdS/CFT,'' JHEP {\bf0410} (2004) 060, 
  \href{http://arxiv.org/abs/hep-th/0410105}{\tt arXiv:\hspace{0pt}hep-th\hspace{0pt}/\hspace{0pt}0410105}
  

\bibitem{BKSZ} N.~Beisert, V.~Kazakov, K.~Sakai, K.~Zarembo, ``The Algebraic
  Curve of Classical Superstrings on $AdS_5\times S^5$,'' Commun.\ Math.\ Phys.\
  {\bf263} (2006) 659, \arxiv{hep-th/0502226}

\bibitem{ALGREVIEW} S.~Schäfer-Nameki, ``Review of AdS/CFT Integrability,
  Chapter II.4: The Spectral Curve,'' Lett.\ Math.\ Phys.\ {\bf99} (2011) 169,
  \arxiv[hep-th]{1012.3989}

\bibitem{AM}
  L.~F.~Alday, J.~Maldacena,
  ``Null polygonal Wilson loops and minimal surfaces in Anti-de-Sitter space,''
  JHEP {\bf 0911} (2009) 082,
  \arxiv[hep-th]{0904.0663}

\bibitem{AGM}
  L.~F.~Alday, D.~Gaiotto, J.~Maldacena,
  ``Thermodynamic Bubble Ansatz,''
  JHEP {\bf 1109} (2011) 032,
  \arxiv[hep-th]{0911.4708}

\bibitem{YW}
  L.~F.~Alday, J.~Maldacena, A.~Sever, P.~Vieira,
  ``Y-system for Scattering Amplitudes,''
  J.\ Phys.\ A: Math.\ Theor.\ {\bf 43} (2010) 485401,
  \arxiv[hep-th]{1002.2459}

\bibitem{JSW}
  R.~A.~Janik, P.~Surówka, A.~Wereszczyński,
  ``On correlation functions of operators dual to classical spinning string states,''
  JHEP {\bf 1005} (2010) 030,
  \arxiv[hep-th]{1002.4613}

\bibitem{BT}
  E.~I.~Buchbinder, A.~A.~Tseytlin,
  ``On semiclassical approximation for correlators of closed string vertex operators in AdS/CFT,''
  JHEP {\bf 1008} (2010) 057,
  \arxiv[hep-th]{1005.4516}

\bibitem{JW}
  R.~A.~Janik, A.~Wereszczyński,
  ``Correlation functions of three heavy operators: The AdS contribution,''
  JHEP {\bf 1112} (2011) 095,
  \arxiv[hep-th]{1109.6262}

\bibitem{BT2}
  E.~I.~Buchbinder, A.~A.~Tseytlin,
  ``Semiclassical correlators of three states with large $S^5$ charges in string theory in $AdS_5\times S^5$,''
  Phys.\ Rev.\ D {\bf 85} (2012) 026001,
  \arxiv[hep-th]{1110.5621}

\bibitem{CORRLOOP} K.~Zarembo, ``Open string fluctuations in $AdS_5\times S^5$
  and operators with large R charge,'' Phys.\ Rev.\ D {\bf66} (2002) 105021,
  \arxiv{hep-th/0209095}

\bibitem{GKP}
  S.~S.~Gubser, I.~R.~Klebanov, A.~M.~Polyakov,
  ``A Semiclassical limit of the gauge/string correspondence,''
  Nucl.\ Phys.\ B {\bf 636} (2002) 99,
  \arxiv{hep-th/0204051}

\bibitem{KRUCZGKP}
  M.~Kruczenski,
  ``A Note on twist two operators in $\mathcal{N}=4$ SYM and Wilson loops in Minkowski signature,''
  JHEP {\bf 0212} (2002) 024
  \arxiv{hep-th/0210115}.

\bibitem{FD}
  N.~Drukker, V.~Forini,
  ``Generalized quark-antiquark potential at weak and strong coupling,''
  JHEP {\bf 1106} (2011) 131,
  \arxiv[hep-th]{1105.5144}

\bibitem{CHMSqq}
  D.~Correa, J.~Maldacena, A.~Sever,
  ``The quark anti-quark potential and the cusp anomalous dimension from a TBA equation,''
  \arxiv[hep-th]{1203.1913}

\bibitem{Drukkerqq}
  N.~Drukker,
  ``Integrable Wilson loops,''
  \arxiv[hep-th]{1203.1617}

\bibitem{BPR}
  I.~Bena, J.~Polchinski, R.~Roiban,
  ``Hidden symmetries of the $AdS_5\times S^5$ superstring,''
  Phys.\ Rev.\ D {\bf 69} (2004) 046002,
  \arxiv{hep-th/0305116}

\bibitem{DV} N.~Dorey, B.~Vicedo, ``On the Dynamics of Finite-Gap Solutions in
  Classical String Theory,'' JHEP {\bf0607} (2006) 014, \arxiv{hep-th/0601194}

\bibitem{BABELON} O.~Babelon, D.~Bernard, M.~Talon, ``Introduction to Classical
  Integrable Systems,'' Cambridge University Press, 2003

\bibitem{DVPHD} B.~Vicedo, ``Finite-g Strings,'' PhD thesis,
  \arxiv[hep-th]{0810.3402}\\
  B.~Vicedo, ``The method of finite-gap integration in classical and
  semi-classical string theory,'' J.\ Phys.\ A: Math.\ Theor.\ {\bf44} (2011) 124002

\bibitem{CUSPSOL} R.~Roiban, A.~Tseytlin, ``Strong-coupling expansion of cusp
  anomaly from quantum superstring,'' JHEP {\bf0711} (2007) 016,
  \arxiv[hep-th]{0709.0681}

\bibitem{QQMALD} J.~Maldacena, ``Wilson loops in large $N$ field theories,''
  Phys.\ Rev.\ Lett.\  {\bf80} (1998) 4859, \arxiv{hep-th/9803002}

\bibitem{SJREY}
  S.-J.~Rey, J.-T.~Yee,
  ``Macroscopic strings as heavy quarks in large $N$ gauge theory and anti-de Sitter supergravity,''
  Eur.\ Phys.\ J.\ C {\bf 22} (2001) 379,
  \arxiv{hep-th/9803001}

\bibitem{QQCHR} S.-x.~Chu, D.~Hou, H.-c.~Ren, ``The Subleading Term of the
  Strong Coupling Expansion of the Heavy-Quark Potential in a $\mathcal{N}=4$ Super
  Yang-Mills Vacuum,'' JHEP {\bf0908} (2009) 004, \arxiv[hep-ph]{0905.1874}

\bibitem{Bobenko}
  M.~Babich, A.~Bobenko, ``Willmore Tori with umbilic lines and minimal surfaces
  in hyperbolic space,'' Duke Mathematical Journal {\bf 72}, No. 1, 151 (1993)

\bibitem{Kruczenski}
  R.~Ishizeki, M.~Kruczenski, S.~Ziama,
  ``Notes on Euclidean Wilson loops and Riemann Theta functions,''
  \arxiv[hep-th]{1104.3567}

\bibitem{ES}
  B.~Eden, M.~Staudacher,
  ``Integrability and transcendentality,''
  J.\ Stat.\ Mech.\  {\bf 0611} (2006) P11014,
  \arxiv{hep-th/0603157}

\bibitem{BES}
  N.~Beisert, B.~Eden, M.~Staudacher,
  ``Transcendentality and Crossing,''
  J.\ Stat.\ Mech.\  {\bf 0701} (2007) P01021,
  \arxiv{hep-th/0610251}

\bibitem{STEGUN} M.~Abramowitz, I.~Stegun, eds., ``Handbook of Mathematical
  Functions With Formulas, Graphs and Mathematical Tables'', Dover Publications,
  New York, 1970

\bibitem{NIST} F.~Olver et al., eds., ``NIST Handbook of Mathematical
  Functions,'' Cambridge University Press, 2010

\end{thebibliography}
\end{document}